\documentclass[nonacm,sigconf]{acmart}
\AtBeginDocument{%
  }

\setcopyright{acmlicensed}
\copyrightyear{2025}
\acmYear{2025}
\acmDOI{XXXXXXX.XXXXXXX}

\usepackage{hyperref}       
\usepackage{url}            
\usepackage{multirow}
\usepackage{graphicx}
\usepackage{tcolorbox}
\usepackage{subcaption}

\usepackage{xcolor}

\usepackage{glossaries}
\newacronym{llm}{LLM}{Large Language Model}
\newacronym{rag}{RAG}{Retrieval Augmented Generation}
\newacronym{nlp}{NLP}{Natural Language Processing}
\newacronym{eis}{EIS}{Environmental Impact Statement}
\newacronym{nepa}{NEPA}{National Environment Policy Act}

\begin{document}


\title{Evaluating the Robustness of Dense Retrievers in Interdisciplinary Domains}
\author{%
  Sarthak Chaturvedi, 
  Anurag Acharya,
  Rounak Meyur,
  Koby Hayashi, \\
  \textbf{Sai Munikoti,
  Sameera Horawalavithana}\\
  Pacific Northwest National Laboratory\\
  Richland, WA 99354
}




\renewcommand{\shortauthors}{Chaturvedi et al.}

\renewenvironment{quote}
  {\small\list{}{\rightmargin=0.25cm \leftmargin=0.25cm}%
   \item\relax}
  {\endlist}

\begin{abstract}
Evaluation benchmark characteristics may distort the true benefits of domain adaptation in retrieval models. This creates misleading assessments that influence deployment decisions in specialized domains. 
We show that two benchmarks with drastically different features such as topic diversity, boundary overlap, and semantic complexity can influence the perceived benefits of fine-tuning. Using environmental regulatory document retrieval as a case study, we fine-tune ColBERTv2 model on Environmental Impact Statements (EIS) from federal agencies. We evaluate these models across two benchmarks with different semantic structures.
Our findings reveal that identical domain adaptation approaches show very different perceived benefits depending on evaluation methodology. On one benchmark, with clearly separated topic boundaries, domain adaptation shows small improvements (maximum 0.61\% NDCG gain). However, on the other benchmark with overlapping semantic structures, the same models demonstrate large improvements (up to 2.22\% NDCG gain), a 3.6-fold difference in the performance benefit. 
We compare these benchmarks through topic diversity metrics, finding that the higher-performing benchmark shows 11\% higher average cosine distances between contexts and 23\% lower silhouette scores, directly contributing to the observed performance difference. 
These results demonstrate that benchmark selection strongly determines assessments of retrieval system effectiveness in specialized domains. 
Evaluation frameworks with well-separated topics regularly underestimate domain adaptation benefits, while those with overlapping semantic boundaries reveal improvements that better reflect real-world regulatory document complexity. 
Our findings have important implications for developing and deploying AI systems for interdisciplinary domains that integrate multiple topics.
\end{abstract}


\received{13 June 2025}

\maketitle

\section{Introduction}
\label{sec:intro}
Traditional evaluation approaches for domain-adapted retrieval models may distort the true benefits of fine-tuning, creating a false sense of confidence in model capabilities when deployed in real-world scenarios. The characteristics of evaluation benchmarks including degree of topic diversity and overlap can strongly affect perceived performance improvements, leading different benchmarks to yield conflicting results about the same model. This evaluation gap is especially concerning for automated systems in high-stakes regulatory domains, where reliable performance assessment is critical for readiness for use.

In this study, we assess the retrieval model performance in the domain of environmental reviews conducted under the National Environment Policy Act (NEPA)\footnote{https://www.epa.gov/nepa}.
\gls*{nepa} stands as a foundational piece of environmental legislation in the United States, requiring federal agencies to consider the environmental impacts of their proposed actions\footnote{https://www.whitehouse.gov/presidential-actions/2025/04/updating-permitting-technology-for-the-21st-century/}.  
We focused on the environmental regulatory documents such as Environmental Impact Statements (EISs) that contain interdisciplinary topics that span across domains such as environmental science, policy, and law for studying the performance of retrieval models. 
Traditional keyword-based information retrieval methods fail to capture the specialized terms and word relationships in these documents; for example, a query about "local wildlife impacts" might miss relevant content discussing "faunal ecosystems" or "biodiversity zones." 

Recent advances in contextual embedding models like ColBERT \cite{khattab2020colbert} and ColBERTv2 \cite{santhanam2021colbertv2} have shown promise for addressing these limitations through detailed matching between queries and documents. 
However, their effectiveness decreases when dealing with domain-specific language and concepts not represented in their training data \cite{807600ef43073cd9c59d4208ee710e90cf14efa8, 590432f953b6ce1b4b36bf66a2ac65eeee567515,ko2025denseretrieversupdatedevolving}. 
While domain adaptation through fine-tuning offers a solution to this challenge, evaluating whether such adaptations truly improve real-world performance depends heavily on the evaluation benchmarks used.
This creates a basic question: \textit{Does better performance on standard benchmarks translate to better performance in varied, difficult scenarios?}

Our work tackles this gap by studying how different benchmark characteristics show different sides of model performance when evaluating domain-adapted retrieval systems. 
We demonstrate how benchmarks with different characteristics can strongly influence our assessment of domain adaptation benefits.
We fine-tune ColBERTv2 models on a growing corpus of EIS documents using synthetic question-context pairs.
To evaluate adaptation effectiveness, we employ two benchmarks which differ significantly in their topic diversity and boundary characteristics. 
We analyze how topic diversity metrics including cosine distance, silhouette score, and topic entropy relate to the model performance improvements, showing the important connection between benchmark complexity and how effective the model appears.
Our findings contribute to the development of more robust evaluation methods for interdisciplinary domains with complex, and overlapping topics and provides a better understanding of when and how domain adaptation benefits emerge, helping create more reliable evaluation methodologies for high-stakes domains.

\section{Related Works}
\label{sec:related}
Domain adaptation for retrieval systems has shown promise in specialized fields, with models like BioBERT for biomedical text mining \cite{lee2020biobert} and LegalBERT for legal documents \cite{chalkidis2020legal} showing better understanding of domain-specific content. However, studies have revealed that while these domain-adapted language models improve contextual understanding, they may not lead to improved retrieval effectiveness without specific adaptation for retrieval tasks \cite{501397e553ce88c2680c287dc18446e7494ee59d, saad2023udapdr}. 

The BEIR benchmark \cite{thakur2021beir} further demonstrates that retrieval models often struggle when applied to new, specialized domains without domain-specific training. Even with this evidence, adapting embedding-based retrieval models like ColBERTv2 to specialized domains has received little attention \cite{Zhong2022ApplyingSA, 10755364, hou2024clercdatasetlegalcase}. 
Importantly, existing work in this area mainly focuses on improving model architectures and training procedures instead of asking if our evaluation methodologies properly show the true benefits of domain adaptation.

Standard evaluation methods for information retrieval systems rely on standard metrics like precision, recall, and NDCG across benchmark datasets \cite{moffat2017incorporating, white2016interactions}, but these approaches may not show real-world performance in specialized domains. 
More recently, Hsia et al. \cite{hsia2024} show how evaluation methods for retrieval-augmented generation can lead to wrong conclusions about system performance when not carefully designed. Studies have shown that inaccuracies in IR systems hinder user adoption, showing the need for better accuracy and reliability \cite{nenkova2010framework, yao2019empirical}. However, the characteristics of evaluation datasets themselves can regularly affect performance assessment, yet this aspect is seldom studied \cite{bf6b3a4cc6666b86854b54d11eb4aaf6968aaf5f}. 
This evaluation gap is especially problematic in high-stakes domains like environmental regulatory compliance~\cite{nepaquad}, legal case retrieval \cite{hou2024clercdatasetlegalcase}, and healthcare, where the gap between standard evaluation practices and real-world use cases can create false confidence in system capabilities and reduce user trust and adoption.

Even though reliable evaluation is critical, most domain adaptation research assumes that benchmark choice does not strongly affect conclusions about model improvement. Few studies carefully study how benchmark characteristics such as topic diversity, complexity, and boundary overlap affect evaluation outcomes and apparent adaptation benefits. While synthetic data generation approaches like UDAPDR \cite{saad2023udapdr} have become good solutions for dealing with limited labeled data in domain adaptation, their effectiveness is usually tested with standard evaluation protocols that may not reflect real-world complexity. 

\section{Methodology}
\label{sec:methods}
To investigate how benchmarks influence the assessment of domain adaptation benefits, we designed a controlled experiment using environmental regulatory document retrieval as our case study. 
Our methodology examines how models fine-tuned with varying levels of domain exposure perform differently across benchmarks with different topic diversity properties. 
In this section, we describe our methodology including the data collection and preprocessing (Section~\ref{subsec:results_data}), synthetic data generation (Section~\ref{subsec:methods_data_generation}), model fine-tuning (Section~\ref{subsec:methods_finetuning}), and benchmark evaluation (Section~\ref{subsec:evaluation}).

\subsection{Data Collection and Preprocessing}
\label{subsec:results_data}

We collected a complete set of over 700 Environmental Impact Statements (EISs) from various federal agencies, including the Department of Energy, Department of Transportation, and the Environmental Protection Agency. 
EIS documents are large and often provided as multiple document versions, which can include appendices, executive summaries, comments, and other additional materials. 
We focused on the complete final versions of the EIS documents needed for good model training.
We filtered the documents to include only the main body of each final EIS, selecting files containing phrases like "Final EIS," "Final Volume," or "Final Vol" in their filenames. 
We excluded files labeled as "Appendix," "Executive Summary," or "Comment," as they usually contain additional materials not central to the main content. 

We extracted text from the final EIS documents and cleaned them by removing any leftover metadata, headers, footers, and formatting inconsistencies to ensure uniformity and readability. 
We used the \texttt{LlamaIndex} sentence splitter to divide the text into sentences.
We grouped sentences into logical chunks of up to 256 tokens without splitting sentences mid-way. 
This process resulted in a structured dataset of manageable text chunks with logical flow and context, which is important for good retrieval.

Given the large number of chunks generated from the documents, processing all of them for synthetic data generation would be very demanding on computing resources. 
We randomly selected 30\% of the chunks from each document for synthetic data generation. 
This sampling was performed per document, while all documents contributed equally to the dataset. 
This strategy resulted a dataset with a diverse set of topics and domains.

\begin{figure*}
  \centering
  \begin{subfigure}[b]{0.46\textwidth}
    \centering
    \includegraphics[width=\textwidth]{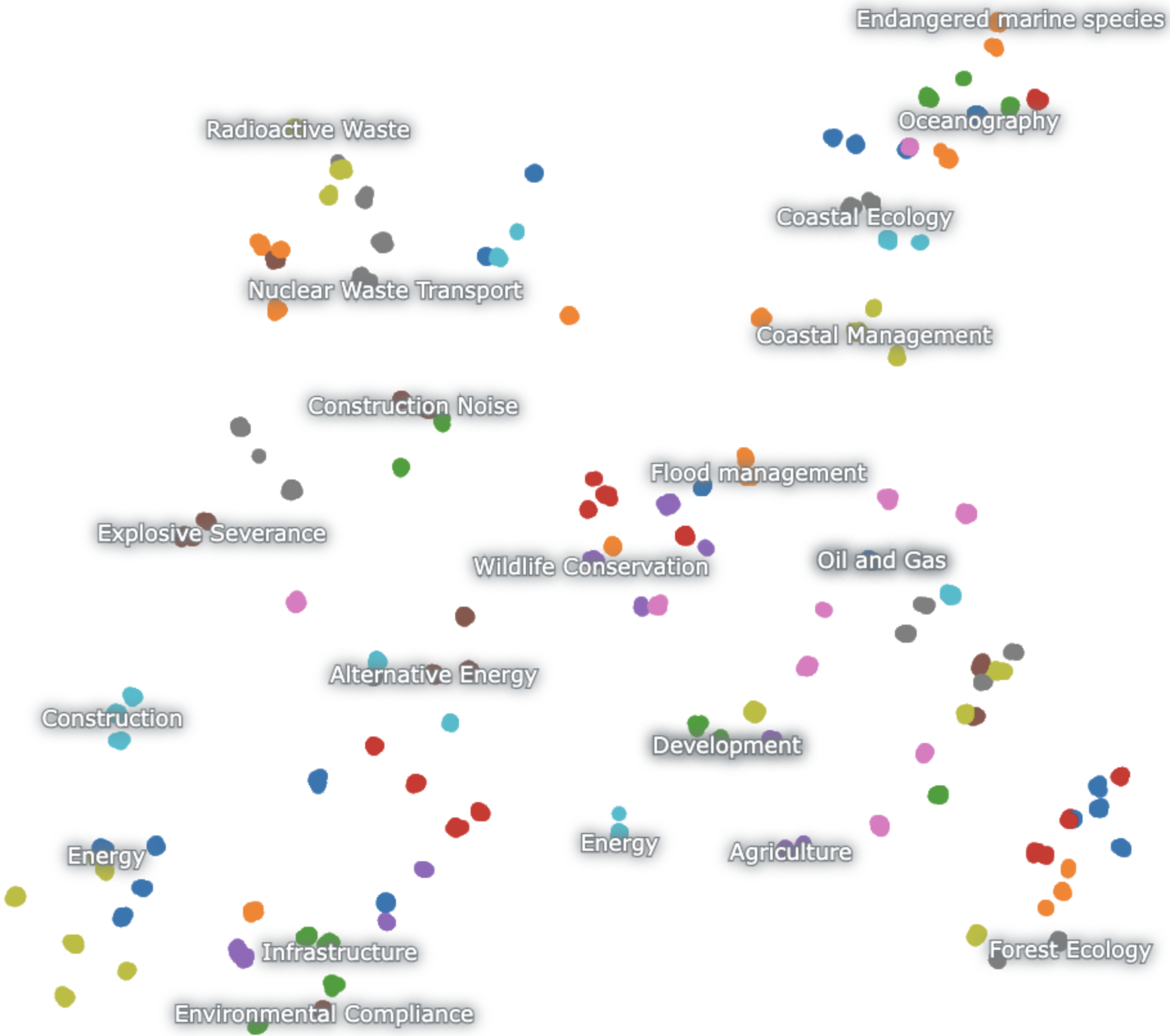}
    \caption{NEPAQuAD-SME-LLM}
    \label{fig:nepa_quad}
  \end{subfigure}
  \hspace{0.01\textwidth}
  %
  \begin{minipage}[b]{0.01\textwidth}
    \centering
    \rule{0pt}{0.6\textwidth} 
    \rotatebox{90}{\dotfill}
  \end{minipage}
  \hspace{0.01\textwidth}
  \begin{subfigure}[b]{0.46\textwidth}
    \centering
    \includegraphics[width=\textwidth]{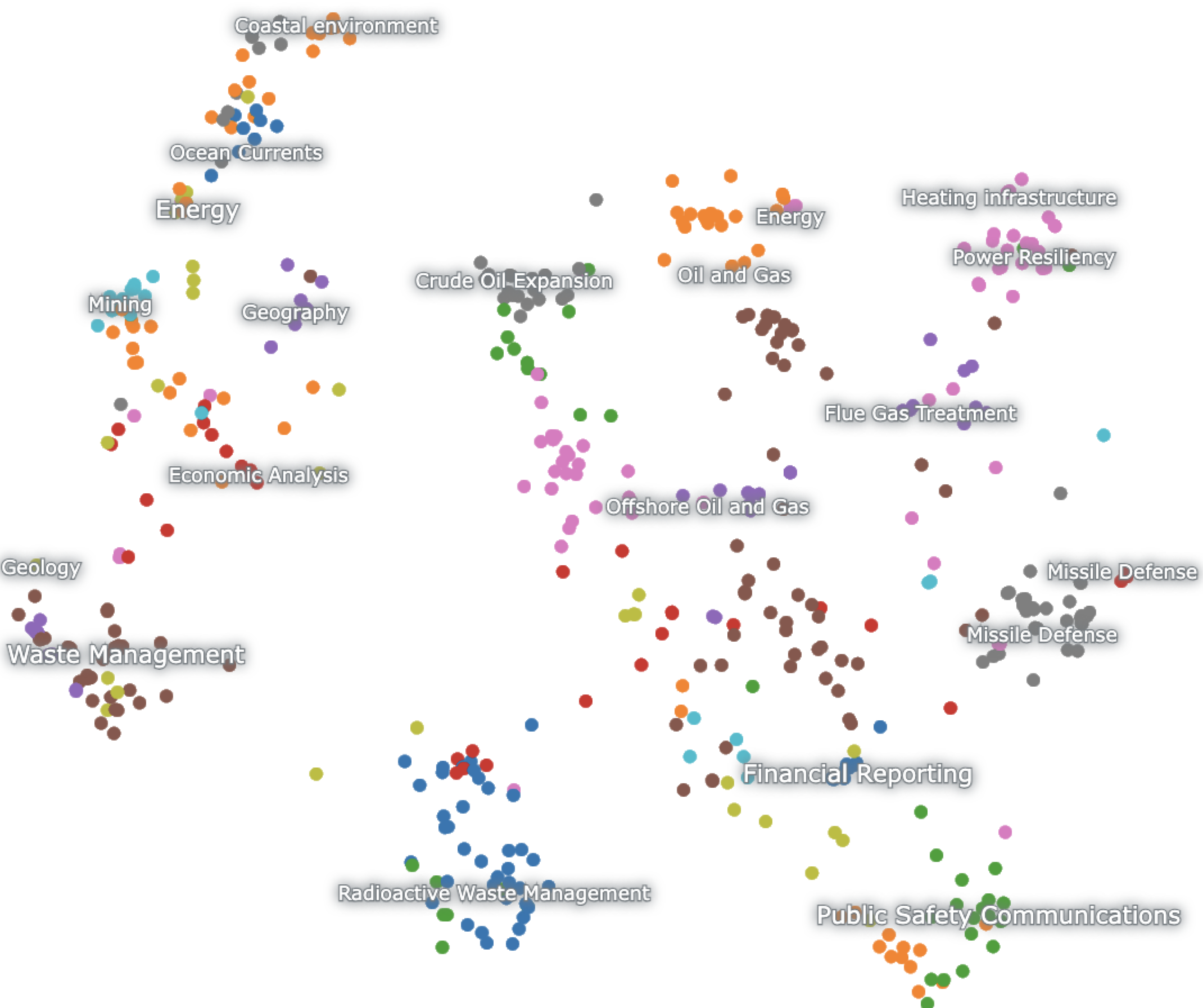}
    \caption{NEPAQuAD-LLM}
    \label{fig:ercts}
  \end{subfigure}
  \caption{Topic boundary visualization comparing NEPAQuAD-SME-LLM (a) and NEPAQuAD-LLM (b). NEPAQuAD-SME-LLM shows distinct clustering while NEPAQuAD-LLM exhibits greater topic overlap and spread out boundaries.}
  \label{fig:topic_boundaries}
\end{figure*}

\subsection{Synthetic Data Generation}
\label{subsec:methods_data_generation}

To create a domain-specific training dataset, we generated synthetic question-context pairs and hard negative contexts.

\subsubsection{\textbf{Generating Query-Answer Pairs}}
For each sampled chunk, we used \texttt{Gemini 1.5 Pro} language model from Google Vertex AI to generate synthetic question-context pairs. The language model was prompted to produce high-quality, relevant questions and their corresponding answers based solely on the information present in the chunk. An example prompt is shown above.

This method created sets of synthetic question-answer pairs $\{(q_j, a_j)\}$ associated with their respective contexts $C_i$, capturing the wide range of potential information needs within the NEPA domain.

\begin{tcolorbox}[colback=gray!5!white, colframe=gray!75!black, title=Prompt]
\textit{You are an expert AI assisting in creating a high-quality, diverse synthetic dataset to train information retrieval models. Analyze the following document chunk and generate potential queries along with their corresponding answers based on the information present. If the context does not contain sufficient information, return empty lists.\\ \textbf{Context:} [Document Chunk]} \\
\end{tcolorbox}

\subsubsection{\textbf{Hard Negative Context Generation}}
\label{sec:hard-negatives}
To improve the model's ability to distinguish between relevant and irrelevant content, we added hard negative contexts into the training data. Hard negatives are chunks that are semantically similar to the positive context but do not contain the correct answer, forcing the model to distinguish between closely related but irrelevant information. For each question–context pair, we used \texttt{RAGatouille}\footnote{\url{https://github.com/abiwang/ragatouille}} to mine hard negative examples that are semantically similar contexts that do not contain the correct answer.

More specifically, for each query \(q_j\), the function retrieved the top 10 contexts \(\{C_{j,k}^-\}_{k=1}^{10}\) from the entire corpus across all documents that were most semantically similar to the query yet distinct from the context containing the correct answer. These hard negatives create challenging training examples where the model must learn to distinguish between truly relevant contexts and those that only contain similar terminology or concepts but lack the specific information needed to answer the query.

\subsubsection{\textbf{Dataset Statistics}}
We generated synthetic question-context pairs for model training using datasets created from 10, 100, and 700 EIS documents, respectively. 
Table~\ref{table:synthetic-data} shows the key statistics for each dataset, including the distribution across EIS documents, agencies who authored these documents, text chunks, and question-context pairs generated. 
Each positive question–context pair was paired with its 10 hard negative contexts mined using the approach described in Section~\ref{sec:hard-negatives}, creating the training triplets used for model fine-tuning.

Scaling from 10 to 700 EIS documents greatly improved not only the data volume but also its diversity with respect to source agencies. The representation of federal agencies increased from 8 to 83, providing the model with exposure to a wider range of terminologies, regulatory frameworks, and domain-specific contexts. While this training data diversity (represented by federal agency coverage) improves model generalization, evaluation benchmark characteristics (topic boundary overlap, semantic diversity) determine how well we can assess this improvement.

\begin{table}
    \small
    \centering
    \caption{Summary of Synthetic Data Generated for Each Training Dataset that span on 10, 100 and 700 EIS documents}
    \label{table:synthetic-data}
    \begin{tabular}{|r|r|r|r|} \hline        
        \textbf{\#Documents} & \textbf{\#Agencies} & \textbf{\#Chunks} & \textbf{Question \& Context \#Pairs} \\ \hline
        10  & 8  & 17,169  & 2,849   \\ \hline
        100 & 33 & 159,848 & 27,986  \\ \hline
        700 & 83 & 953,440 & 761,980 \\ \hline
    \end{tabular}
\end{table}

\subsection{Model Fine-Tuning}
\label{subsec:methods_finetuning}

Given the limitations of general-purpose embedding models in capturing the specialized language of NEPA documents (see Appendix~\ref{appendix:preliminary-experiments}), we selected ColBERTv2 for adaptation due to its ability to perform detailed token-level interactions and its strong performance in understanding context. To help with the fine-tuning process, we used the \texttt{RAGatouille}, a specialized training framework designed for fine tuning ColBERT.

To investigate whether benchmark characteristics affect evaluation differently depending on the stage of domain adaptation, we conducted step-by-step fine-tuning experiments with ColBERTv2 using synthetic datasets of varying scales. Models with limited domain exposure (10 EIS) may show different sensitivity to topic boundary overlap compared to heavily adapted models (700 EIS), allowing us to determine if benchmark effects are consistent across adaptation levels or vary with training data scale:

\begin{itemize}
\item \textbf{Early Adaptation Stage (10 EIS Documents)}: Limited domain-specific exposure to assess benchmark sensitivity with minimal NEPA-specific training.
\item \textbf{Intermediate Adaptation Stage (100 EIS Documents)}: Moderate domain exposure with greater agency diversity to examine benchmark effects at mid-adaptation levels.
\item \textbf{Advanced Adaptation Stage (700 EIS Documents)}: Detailed domain-specific content to evaluate how benchmark characteristics affect assessment of fully adapted models.
\end{itemize}

This step-by-step approach allows us to examine whether different benchmark characteristics consistently influence evaluation across adaptation stages, providing insights into the basic relationship between benchmark properties and perceived model effectiveness. For each fine-tuning experiment, we maintained consistency in the training procedure (see Appendix~\ref{appendix:training-procedure}) to ensure fair comparison across different adaptation stages.

\subsection{Evaluation}
\label{subsec:evaluation}

Our evaluation methodology provides a framework for understanding how different benchmark characteristics impact our assessment of domain adaptation effectiveness.

\subsubsection{\textbf{Evaluation Datasets}}
We used two different test sets with different characteristics to examine how benchmark properties influence evaluation outcomes:

\begin{itemize}
\item \textbf{NEPAQuAD-SME-LLM} (NQ-SME-LLM): A focused benchmark containing 1589 question-context pairs with 89 unique contexts, representing more clearly separated information needs with fairly distinct topic boundaries \cite{nepaquad}. Both LLM and SME inputs were used to create this benchmark.
\item \textbf{NEPAQuAD-LLM} (NQ-LLM): A complete benchmark with 556 question-context pairs and 507 unique contexts, representing a more challenging and realistic retrieval scenario with significant topic boundary overlap. This set is generated by LLM without SME input. 
\end{itemize}

Both datasets were created by selecting Environmental Impact Statement (EIS) documents that do not overlap with our training set. We used the \texttt{Gemini 1.5 Pro} language model to generate high-quality, relevant synthetic question-context pairs across six different types of questions: inference, closed-ended, comparison, process, divergent, and evaluation. Please refer to Appendix~\ref{appendix:test-dataset} for more details on the evaluation datasets.

\subsubsection{\textbf{Topic Diversity Analysis}}
To measure the differences between our evaluation datasets and understand how benchmark characteristics might influence assessment, we conducted a detailed analysis of their topic diversity properties as shown in Table~\ref{table:diversity_metrics}.

\begin{table}
    \centering
    \caption{Topic diversity metrics across evaluation benchmarks}
    \label{table:diversity_metrics}
    \begin{tabular}{|l|c|c|c|}
    \hline
        \textbf{Metric} & \textbf{NQ-SME-LLM} & \textbf{NQ-LLM} & \textbf{Difference} \\
    \hline
        Avg. Cosine Distance & 0.2321 & 0.2579 & +11.1\% \\
    \hline
        Optimal \# of Clusters & 20 & 19 & -5.0\% \\
    \hline
        Silhouette Score & 0.1030 & 0.0791 & -23.2\% \\
    \hline
        Topic Entropy & 0.9577 & 0.9765 & +2.0\% \\
    \hline
    \end{tabular}
\end{table}

This analysis revealed that NEPAQuAD-LLM has much higher average cosine distances between contexts, indicating greater semantic diversity. Also, NEPAQuAD-LLM shows a lower silhouette score, suggesting less clearly defined topic boundaries and a higher degree of topic overlap. These different characteristics allow us to examine how benchmark properties regularly influence the perceived benefits of domain adaptation.

\subsubsection{\textbf{Evaluation Method}}
All documents are processed into a index using \texttt{Langchain} and \texttt{RAGatouille} \cite{Clavié2025AnswerDotAI}. Some contexts that exceeded 512 tokens were shortened to fit the embedding models maximum token size.
For each question in the test datasets, we used the retrieval models to rank all available contexts from the indexed EIS documents. A context was considered relevant if it contained the information needed to answer the question. For each question, we had a single gold-standard context (the original context from which the question was created), which served as the ground truth for relevance assessment.

\subsubsection{\textbf{Evaluation Metrics}}
We evaluate model performance using Normalized Discounted Cumulative Gain (NDCG) \cite{jarvelin2002cumulated}, which measures how well the retrieval model ranks relevant contexts, with higher weights assigned to relevant contexts appearing at higher positions.
Since the gold-standard contexts were used to generate the questions they are assigned a relevance of $1$ while all other contexts are assigned a relevance of $0$.
For the NEPAQuAD-SME-LLM benchmark, we report NDCG@89, and for the NEPAQuAD-LLM benchmark, we report NDCG@507, corresponding to the total number of unique contexts that we rank in each dataset.

\section{Results}
\label{sec:results}

\begin{figure*}
  \centering
  \includegraphics[width=0.95\textwidth]{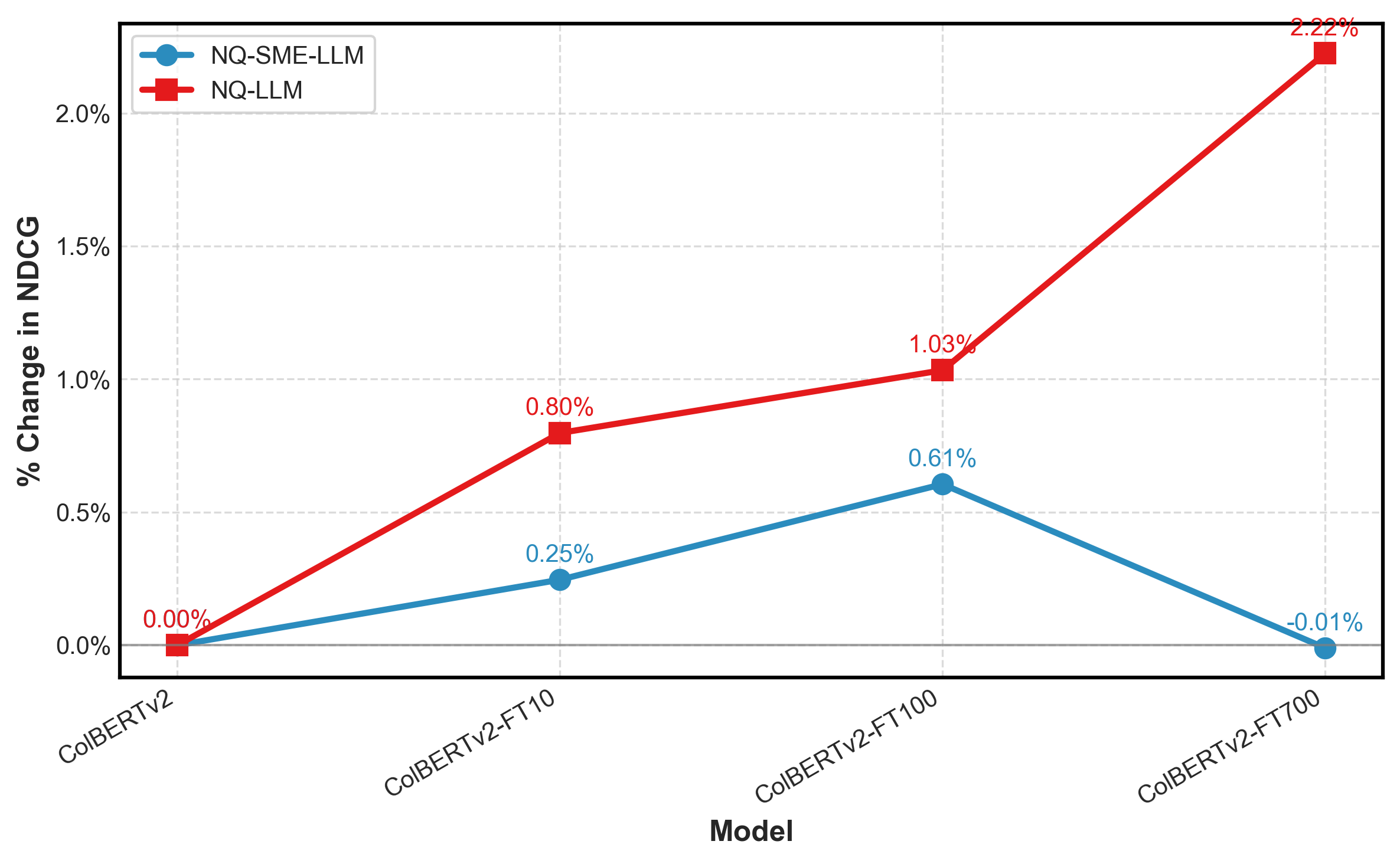}
  \caption{NDCG improvement comparison across benchmarks. NEPAQuAD-LLM ($k=509$) demonstrates greater model differentiation with clear domain adaptation benefits, while NEPAQuAD-SME-LLM $(k=89)$ exhibits minimal performance differences, illustrating how benchmark characteristics influence adaptation assessment.}
  \label{fig:ndcg_percentage}
\end{figure*}

Our study reveals how benchmark characteristics can strongly affect the assessment of domain adaptation benefits in retrieval models. Through careful comparison of models across benchmarks with different topic structures, we demonstrate that the perceived value of domain adaptation varies greatly depending on evaluation methodology.

\subsection{Benchmark Characteristics Create Distinct Evaluation Contexts}

Our analysis reveals major differences between the NEPAQuAD-SME-LLM and NEPAQuAD-LLM evaluation benchmarks (Table~\ref{table:diversity_metrics}). NEPAQuAD-LLM presents a much more challenging retrieval environment with 11.1\% greater average semantic distance between contexts and 23.2\% lower silhouette scores than NEPAQuAD-SME-LLM. This indicates that NEPAQuAD-LLM shows less clearly defined topic boundaries and greater semantic overlap between contexts. The higher topic entropy in NEPAQuAD-LLM (2.0\% increase) confirms greater topic distribution complexity, creating scenarios where models must use better abilities to distinguish between semantically similar content.

The visualization in Figure~\ref{fig:topic_boundaries} supports these findings, showing distinct clustering patterns in NEPAQuAD-SME-LLM compared to the overlapping, spread out topic structure in NEPAQuAD-LLM. These different characteristics create very different retrieval challenges despite both benchmarks containing regulatory content from similar sources.

\begin{figure*}
  \centering
  \includegraphics[width=0.95\textwidth]{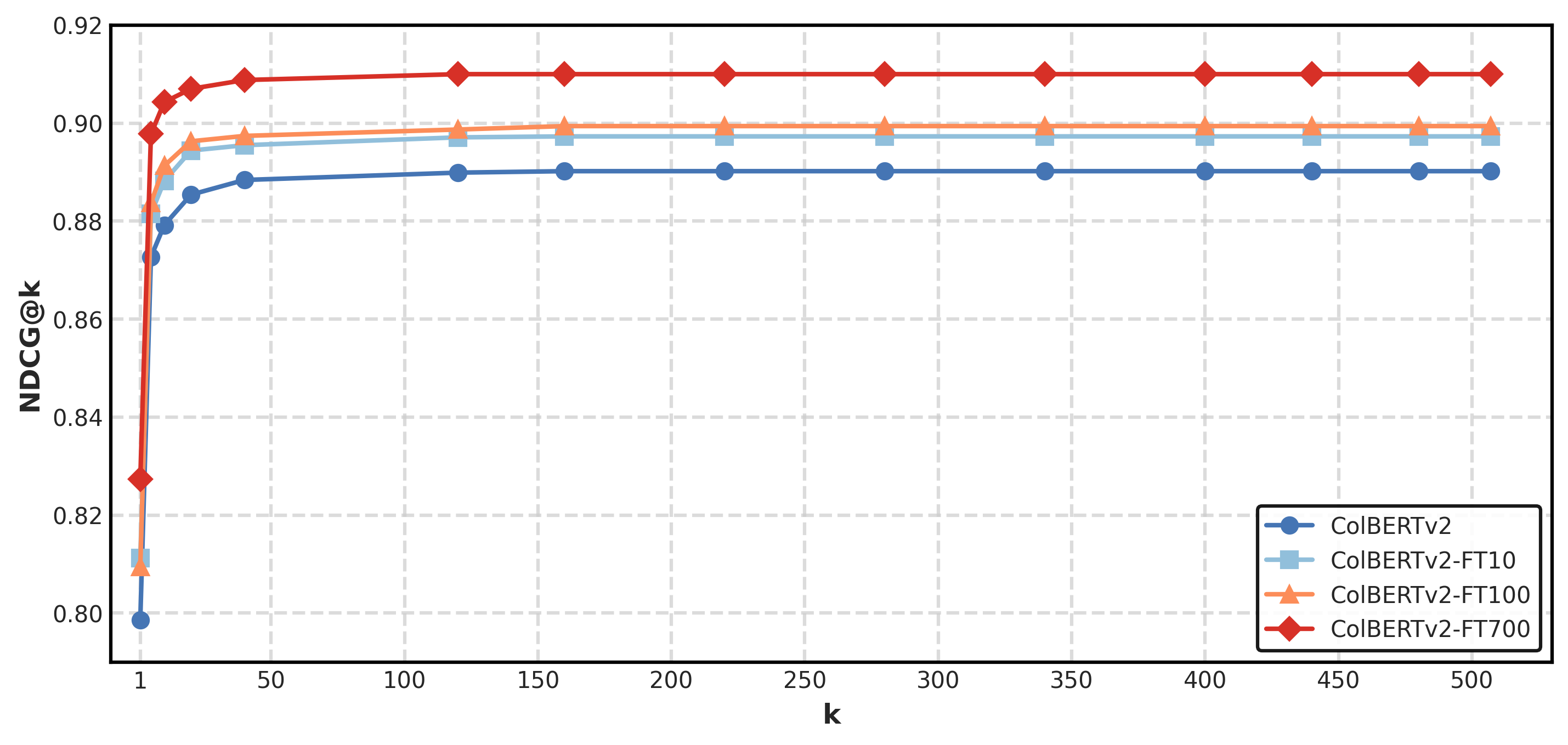}
  \caption{NDCG@k performance comparison of ColBERTv2 models on NEPAQuAD-LLM. All models show performance improvements at lower k values before plateauing, with ColBERTv2-FT700 consistently outperforming other variants at top ranks.}
  \label{fig:ndcg_comparison}
\end{figure*}

\subsection{Performance Improvements Vary Greatly Across Benchmarks}

The size of domain adaptation improvements differs greatly between the two evaluation contexts (Table~\ref{table:ndcg_performance}). On NEPAQuAD-SME-LLM, with its clearly separated semantic clusters, all models achieved high performance (NDCG@5 $>$ 0.97), with small differences from domain adaptation (maximum improvement of +0.61\% with ColBERTv2-FT100). The high baseline performance and small improvement margins indicate limited sensitivity to adaptation effects.

NEPAQuAD-LLM presents very different performance patterns. The evaluation framework with overlapping semantic structures reveals a clear progression of performance improvements with increased domain adaptation. The most heavily fine-tuned model (ColBERTv2-FT700) achieved a 2.22\% improvement over the baseline—3.6 times greater than the maximum improvement observed on NEPAQuAD-SME-LLM. These results demonstrate consistent patterns across multiple experimental runs.


\begin{table}
    \centering
    \caption{Model performance comparison across evaluation benchmarks using NDCG}
    \label{table:ndcg_performance}
    \begin{tabular}{|l|c|c|}
    \hline
        \textbf{Model} & \textbf{NQ-SME-LLM (NDCG)} & \textbf{NQ-LLM (NDCG)} \\
    \hline
        ColBERTv2 & 0.9749 & 0.8902 \\
    \hline
        ColBERTv2-FT10 & 0.9773 & 0.8973 \\
    \hline
        ColBERTv2-FT100 & 0.9808 & 0.8994 \\
    \hline
        ColBERTv2-FT700 & 0.9748 & 0.9100 \\
    \hline
    \end{tabular}
\end{table}

Importantly, the 23.2\% lower silhouette score in NEPAQuAD-LLM relates to 3.6 times greater adaptation benefits observed (2.22\% vs 0.61\% maximum improvement). This relationship suggests that benchmark topic boundary clarity relates to perceived adaptation effectiveness, though we note this observation is based on our comparison of these two benchmarks. Evaluation benchmarks with well-separated topics may regularly underestimate the value of domain adaptation for real-world applications.

Detailed analysis of ranking quality at different cutoff points (Figure~\ref{fig:ndcg_percentage} and Figure~\ref{fig:ndcg_comparison}) supports these findings, showing that domain-adapted models consistently outperform baseline models across all ranking positions on NEPAQuAD-LLM. The ColBERTv2-FT700 model performs better and shows clear improvements in top ranks at 1-10, which are important for practical applications where top-ranked results determine user decision-making effectiveness.

\subsection{Source Similarity Controls for Content Variations}

Both evaluation benchmarks come from Environmental Impact Statements from similar federal agencies (see Appendix~\ref{appendix:agency_distribution} for complete agency breakdown), yet create very different retrieval challenges due to their different topic structures. The agency distribution analysis reveals similar representation across both benchmarks, with approximately uniform distribution across agency sources, showing clearly that performance differences stem from topic boundary characteristics rather than document source variations or agency-specific terminology differences.


This source similarity is important for isolating the impact of benchmark characteristics on evaluation outcomes. Both benchmarks require models to distinguish between semantically similar regulatory terminology, but this discrimination task becomes much more challenging in NEPAQuAD-LLM where topics show greater overlap. The benefits of domain adaptation become most clear in these complex scenarios where fine-tuned models use specialized domain knowledge to distinguish between similar content with different informational value.


\section{Conclusion}
\label{sec:conclusion}

Our research question asked: \textit{How do benchmark characteristics (topic diversity, boundary overlap, complexity) influence our assessment of domain adaptation benefits, and what does this reveal about designing more reliable evaluation frameworks for specialized domains?} 
Our findings provide a clear answer: evaluation methodologies strongly determine how we perceive the value of domain adaptation in retrieval models, with important implications for both research and practice.

When evaluated on the NEPAQuAD-SME-LLM benchmark with distinct topic boundaries, all models performed very well with small differences between them (maximum 0.61\% improvement). However, on the NEPAQuAD-LLM benchmark with overlapping semantic structures, domain adaptation showed large performance improvements of up to 2.22\% in NDCG. This is a 3.6 times greater benefits that remained completely hidden in the simpler evaluation context. 
This difference demonstrates that evaluation methodology can make the same domain adaptation approach appear either barely helpful or highly useful, showing an important problem in current evaluation practices.
These different results highlight a basic methodological insight: adaptation benefits are not consistent and depend greatly on the complexity of evaluation contexts.

The implications for environmental regulatory compliance governed by NEPA are important and varied. Distinguishing between concepts like "habitat restoration" versus "habitat mitigation" or "direct impacts" versus "cumulative impacts" can determine regulatory compliance outcomes, where the 2.22\% performance improvement represents the difference between identifying or missing important regulatory requirements. Using evaluation approaches with clearly separated topics may lead to regular underestimation of adaptation benefits and possibly underinvestment in specialized model development for regulatory contexts. This could result in deployment of poorly adapted systems, leading to incomplete environmental assessments, poor stakeholder consultation, project delays, and poor environmental protection.

Future research should develop standard evaluation methods that carefully change complexity across multiple levels, creating evaluation sets that capture the full spectrum of real-world retrieval challenges. Also, research should investigate whether our findings apply to other specialized domains such as legal case retrieval, medical diagnosis support, and financial compliance, where similar complexity and high-stakes decision-making requirements exist. Testing across domains would show whether the relationship between evaluation complexity and perceived adaptation benefits represents a common rule for AI system assessment.

\section{Limitations}
\label{sec:limitations}
Despite the meaningful insights provided by our study, several limitations should be discussed. First, our evaluation benchmarks, while carefully constructed, use synthetic questions generated by large language models. Though we implemented careful quality checks, these questions may not fully capture the specific information needs of actual regulatory practitioners. Future work should confirm our findings using human generated queries from environmental policy experts.

Second, while we showed the relationship between topic boundary characteristics and domain adaptation benefits, we examined only two benchmarks with different structure properties. A more careful study across benchmarks with slowly changing topic overlap characteristics would provide more detailed insights into this relationship. Also, our focus on NEPA and EIS limits the direct applicability of our findings to other regulatory domains, though we expect the core insight about evaluation benchmark characteristics to apply broadly.

Third, our evaluation used mainly NDCG as the evaluation metric. While NDCG captures well ranking quality, it may not show all other important aspects of retrieval system performance in regulatory contexts, such as finding multiple relevant regulatory rules or finding conflicting rules. Developing more specialized evaluation measures that account for these specialized needs could provide more insights into domain adaptation effectiveness.

Finally, computing limits limited our ability to explore bigger domain adaptation or to experiment with more model types. As the field advances, investigating how our findings extend to other retrieval model architectures and larger adaptation scales would better confirm and improve our understanding of how evaluation benchmark features affect domain adaptation assessment.

\section*{Acknowledgement}
\label{sec:ack}
This work was supported by the Office of Policy, U.S. Department of Energy, and Pacific Northwest National Laboratory, which is operated by Battelle Memorial Institute for the U.S. Department of Energy under Contract DE-AC05–76RLO1830.  This paper has been cleared by PNNL for public release as PNNL-SA-212419.

\bibliographystyle{ACM-Reference-Format}
\bibliography{custom}


\begin{thebibliography}{22}


\ifx \showCODEN    \undefined \def \showCODEN     #1{\unskip}     \fi
\ifx \showISBNx    \undefined \def \showISBNx     #1{\unskip}     \fi
\ifx \showISBNxiii \undefined \def \showISBNxiii  #1{\unskip}     \fi
\ifx \showISSN     \undefined \def \showISSN      #1{\unskip}     \fi
\ifx \showLCCN     \undefined \def \showLCCN      #1{\unskip}     \fi
\ifx \shownote     \undefined \def \shownote      #1{#1}          \fi
\ifx \showarticletitle \undefined \def \showarticletitle #1{#1}   \fi
\ifx \showURL      \undefined \def \showURL       {\relax}        \fi
\providecommand\bibfield[2]{#2}
\providecommand\bibinfo[2]{#2}
\providecommand\natexlab[1]{#1}
\providecommand\showeprint[2][]{arXiv:#2}

\bibitem[Alkan et~al\mbox{.}(2024)]%
        {10755364}
\bibfield{author}{\bibinfo{person}{Berkin Alkan},
  \bibinfo{person}{Bekir~Bilgehan Tekin}, \bibinfo{person}{Alper
  Karamanlıoğlu}, {and} \bibinfo{person}{İsmail Karakaya}.}
  \bibinfo{year}{2024}\natexlab{}.
\newblock \showarticletitle{Analysis of Retrieval Performance for Methods
  Fine-Tuned with ColBERT Architecture}. In \bibinfo{booktitle}{\emph{2024
  Medical Technologies Congress (TIPTEKNO)}}. \bibinfo{pages}{1--4}.
\newblock
\href{https://doi.org/10.1109/TIPTEKNO63488.2024.10755364}{doi:\nolinkurl{10.1109/TIPTEKNO63488.2024.10755364}}


\bibitem[Chalkidis et~al\mbox{.}(2020)]%
        {chalkidis2020legal}
\bibfield{author}{\bibinfo{person}{Ilias Chalkidis}, \bibinfo{person}{M.
  Fergadiotis}, \bibinfo{person}{P. Malakasiotis}, \bibinfo{person}{N.
  Aletras}, {and} \bibinfo{person}{I. Androutsopoulos}.}
  \bibinfo{year}{2020}\natexlab{}.
\newblock \bibinfo{title}{LEGAL-BERT: The Muppets Straight Out of Law School}.
\newblock


\bibitem[Clavié et~al\mbox{.}(2025)]%
        {Clavié2025AnswerDotAI}
\bibfield{author}{\bibinfo{person}{Benjamin Clavié}, \bibinfo{person}{Omar
  Khattab}, \bibinfo{person}{Harrison Chase}, \bibinfo{person}{Anirudh
  Dharmarajan}, \bibinfo{person}{Josh Purtell}, \bibinfo{person}{Minh Nguyen},
  \bibinfo{person}{PrimoUomo89}, \bibinfo{person}{tm17 abcgen},
  \bibinfo{person}{Diego~Peláez Paquico}, \bibinfo{person}{Johannes Aalto},
  \bibinfo{person}{Patrick}, \bibinfo{person}{Peter Goldstein},
  \bibinfo{person}{Shaurya Rohatgi}, \bibinfo{person}{Théo Q.},
  \bibinfo{person}{Vishal Bakshi}, \bibinfo{person}{corrius},
  \bibinfo{person}{mauryaland}, \bibinfo{person}{Sami},
  \bibinfo{person}{Jan~Luca Scheerer}, \bibinfo{person}{James},
  \bibinfo{person}{Géraud Bourdin}, \bibinfo{person}{German Martin},
  \bibinfo{person}{Gautam}, \bibinfo{person}{Deven Mistry},
  \bibinfo{person}{Dale Hille}, {and} \bibinfo{person}{Alex Perez}.}
  \bibinfo{year}{2025}\natexlab{}.
\newblock \bibinfo{title}{RAGatouille}.
\newblock
\urldef\tempurl%
\url{https://github.com/AnswerDotAI/RAGatouille}
\showURL{%
\tempurl}


\bibitem[Hou et~al\mbox{.}(2024)]%
        {hou2024clercdatasetlegalcase}
\bibfield{author}{\bibinfo{person}{Abe~Bohan Hou}, \bibinfo{person}{Orion
  Weller}, \bibinfo{person}{Guanghui Qin}, \bibinfo{person}{Eugene Yang},
  \bibinfo{person}{Dawn Lawrie}, \bibinfo{person}{Nils Holzenberger},
  \bibinfo{person}{Andrew Blair-Stanek}, {and} \bibinfo{person}{Benjamin~Van
  Durme}.} \bibinfo{year}{2024}\natexlab{}.
\newblock \bibinfo{title}{CLERC: A Dataset for Legal Case Retrieval and
  Retrieval-Augmented Analysis Generation}.
\newblock
\showeprint[arxiv]{2406.17186}~[cs.CL]
\urldef\tempurl%
\url{https://arxiv.org/abs/2406.17186}
\showURL{%
\tempurl}


\bibitem[Hsia et~al\mbox{.}(2024)]%
        {hsia2024}
\bibfield{author}{\bibinfo{person}{Jennifer Hsia}, \bibinfo{person}{Afreen
  Shaikh}, \bibinfo{person}{Zhiruo. Wang}, {and} \bibinfo{person}{Graham
  Neubig}.} \bibinfo{year}{2024}\natexlab{}.
\newblock \showarticletitle{RAGGED: Towards Informed Design of Retrieval
  Augmented Generation Systems}. In \bibinfo{booktitle}{\emph{NeurIPS Workshop
  on Adaptive Foundation Models}}.
\newblock


\bibitem[J{\"a}rvelin and Kek{\"a}l{\"a}inen(2002)]%
        {jarvelin2002cumulated}
\bibfield{author}{\bibinfo{person}{K. J{\"a}rvelin} {and} \bibinfo{person}{J.
  Kek{\"a}l{\"a}inen}.} \bibinfo{year}{2002}\natexlab{}.
\newblock \showarticletitle{Cumulated gain-based evaluation of IR techniques}.
\newblock \bibinfo{journal}{\emph{ACM Transactions on Information Systems
  (TOIS)}} \bibinfo{volume}{20}, \bibinfo{number}{4} (\bibinfo{year}{2002}),
  \bibinfo{pages}{422--446}.
\newblock
\urldef\tempurl%
\url{http://scholar.google.de/scholar.bib?q=info:6Bdw8cs-UYMJ:scholar.google.com/&output=citation&hl=de&as_sdt=0,5&ct=citation&cd=0}
\showURL{%
\tempurl}


\bibitem[Khattab and Zaharia(2020)]%
        {khattab2020colbert}
\bibfield{author}{\bibinfo{person}{Omar Khattab} {and} \bibinfo{person}{Matei
  Zaharia}.} \bibinfo{year}{2020}\natexlab{}.
\newblock \showarticletitle{ColBERT: Efficient and Effective Passage Search via
  Contextualized Late Interaction over BERT}. In
  \bibinfo{booktitle}{\emph{Proceedings of the 43rd International ACM SIGIR
  Conference on Research and Development in Information Retrieval}}.
  \bibinfo{pages}{39--48}.
\newblock


\bibitem[Ko et~al\mbox{.}(2025)]%
        {ko2025denseretrieversupdatedevolving}
\bibfield{author}{\bibinfo{person}{Dayoon Ko}, \bibinfo{person}{Jinyoung Kim},
  \bibinfo{person}{Sohyeon Kim}, \bibinfo{person}{Jinhyuk Kim},
  \bibinfo{person}{Jaehoon Lee}, \bibinfo{person}{Seonghak Song},
  \bibinfo{person}{Minyoung Lee}, {and} \bibinfo{person}{Gunhee Kim}.}
  \bibinfo{year}{2025}\natexlab{}.
\newblock \bibinfo{title}{When Should Dense Retrievers Be Updated in Evolving
  Corpora? Detecting Out-of-Distribution Corpora Using GradNormIR}.
\newblock
\showeprint[arxiv]{2506.01877}~[cs.IR]
\urldef\tempurl%
\url{https://arxiv.org/abs/2506.01877}
\showURL{%
\tempurl}


\bibitem[Lee et~al\mbox{.}(2020)]%
        {lee2020biobert}
\bibfield{author}{\bibinfo{person}{Jinhyuk Lee}, \bibinfo{person}{Wonjin Yoon},
  \bibinfo{person}{Sungdong Kim}, \bibinfo{person}{Donghyeon Kim},
  \bibinfo{person}{Sunkyu Kim}, \bibinfo{person}{Chang-Hwan So}, {and}
  \bibinfo{person}{Jaewoo Kang}.} \bibinfo{year}{2020}\natexlab{}.
\newblock \showarticletitle{BioBERT: a Pre-trained Biomedical Language
  Representation Model for Biomedical Text Mining}.
\newblock \bibinfo{journal}{\emph{Bioinformatics}} \bibinfo{volume}{36},
  \bibinfo{number}{4} (\bibinfo{year}{2020}), \bibinfo{pages}{1234--1240}.
\newblock


\bibitem[Moffat et~al\mbox{.}(2017)]%
        {moffat2017incorporating}
\bibfield{author}{\bibinfo{person}{Alistair Moffat}, \bibinfo{person}{Paul
  Bailey}, \bibinfo{person}{Fritz Scholer}, {and} \bibinfo{person}{Peter
  Thomas}.} \bibinfo{year}{2017}\natexlab{}.
\newblock \showarticletitle{Incorporating User Expectations and Behavior into
  the Measurement of Search Effectiveness}.
\newblock \bibinfo{journal}{\emph{ACM Transactions on Information Systems}}
  \bibinfo{volume}{35}, \bibinfo{number}{3} (\bibinfo{year}{2017}),
  \bibinfo{pages}{1--38}.
\newblock


\bibitem[Nenkova et~al\mbox{.}(2010)]%
        {nenkova2010framework}
\bibfield{author}{\bibinfo{person}{Ani Nenkova}, \bibinfo{person}{Kathleen
  McKeown}, \bibinfo{person}{Cristina Ros{\'e}}, {and} \bibinfo{person}{Julia
  Hirschberg}.} \bibinfo{year}{2010}\natexlab{}.
\newblock \showarticletitle{A Framework for Assessing Information Quality and
  Trustworthiness of Digital Information Sources}. In
  \bibinfo{booktitle}{\emph{Proceedings of the 48th Annual Meeting of the
  Association for Computational Linguistics}}. \bibinfo{pages}{51--55}.
\newblock


\bibitem[Phan et~al\mbox{.}(2024)]%
        {nepaquad}
\bibfield{author}{\bibinfo{person}{Hung Phan}, \bibinfo{person}{Anurag
  Acharya}, \bibinfo{person}{Rounak Meyur}, \bibinfo{person}{Sarthak
  Chaturvedi}, \bibinfo{person}{Shivam Sharma}, \bibinfo{person}{Mike Parker},
  \bibinfo{person}{Dan Nally}, \bibinfo{person}{Ali Jannesari},
  \bibinfo{person}{Karl Pazdernik}, \bibinfo{person}{Mahantesh Halappanavar},
  \bibinfo{person}{Sai Munikoti}, {and} \bibinfo{person}{Sameera
  Horawalavithana}.} \bibinfo{year}{2024}\natexlab{}.
\newblock \bibinfo{title}{Examining Long-Context Large Language Models for
  Environmental Review Document Comprehension}.
\newblock
\showeprint[arxiv]{2407.07321}~[cs.CL]
\urldef\tempurl%
\url{https://arxiv.org/abs/2407.07321}
\showURL{%
\tempurl}


\bibitem[Rashidi et~al\mbox{.}(2021)]%
        {bf6b3a4cc6666b86854b54d11eb4aaf6968aaf5f}
\bibfield{author}{\bibinfo{person}{Lida Rashidi}, \bibinfo{person}{J. Zobel},
  {and} \bibinfo{person}{Alistair Moffat}.} \bibinfo{year}{2021}\natexlab{}.
\newblock \bibinfo{title}{Evaluating the Predictivity of IR Experiments}.
\newblock
\href{https://doi.org/10.1145/3404835.3463040}{doi:\nolinkurl{10.1145/3404835.3463040}}


\bibitem[Saad-Falcon et~al\mbox{.}(2023)]%
        {saad2023udapdr}
\bibfield{author}{\bibinfo{person}{Jon Saad-Falcon}, \bibinfo{person}{Omar
  Khattab}, \bibinfo{person}{Keshav Santhanam}, \bibinfo{person}{Radu Florian},
  \bibinfo{person}{Martin Franz}, \bibinfo{person}{Salim Roukos},
  \bibinfo{person}{Avirup Sil}, \bibinfo{person}{Md~Arafat Sultan}, {and}
  \bibinfo{person}{Christopher Potts}.} \bibinfo{year}{2023}\natexlab{}.
\newblock \showarticletitle{UDAPDR: unsupervised domain adaptation via LLM
  prompting and distillation of rerankers}.
\newblock \bibinfo{journal}{\emph{arXiv preprint arXiv:2303.00807}}
  (\bibinfo{year}{2023}).
\newblock


\bibitem[Santhanam et~al\mbox{.}(2021a)]%
        {santhanam2021colbertv2}
\bibfield{author}{\bibinfo{person}{K. Santhanam}, \bibinfo{person}{Omar
  Khattab}, {and} \bibinfo{person}{Christopher R{\'e}}.}
  \bibinfo{year}{2021}\natexlab{a}.
\newblock \showarticletitle{ColBERTv2: Effective and Efficient Retrieval via
  Lightweight Late Interaction}.
\newblock \bibinfo{journal}{\emph{arXiv preprint arXiv:2112.01488}}
  (\bibinfo{year}{2021}).
\newblock


\bibitem[Santhanam et~al\mbox{.}(2021b)]%
        {590432f953b6ce1b4b36bf66a2ac65eeee567515}
\bibfield{author}{\bibinfo{person}{Keshav Santhanam}, \bibinfo{person}{O.
  Khattab}, \bibinfo{person}{Jon Saad-Falcon}, \bibinfo{person}{Christopher
  Potts}, {and} \bibinfo{person}{M. Zaharia}.}
  \bibinfo{year}{2021}\natexlab{b}.
\newblock \bibinfo{title}{ColBERTv2: Effective and Efficient Retrieval via
  Lightweight Late Interaction}.
\newblock \bibinfo{numpages}{3715-3734}~pages.
\newblock
\href{https://doi.org/10.18653/v1/2022.naacl-main.272}{doi:\nolinkurl{10.18653/v1/2022.naacl-main.272}}


\bibitem[Thakur et~al\mbox{.}(2021a)]%
        {thakur2021beir}
\bibfield{author}{\bibinfo{person}{Nandan Thakur}, \bibinfo{person}{Nils
  Reimers}, \bibinfo{person}{Andreas R{\"u}ckl{\'e}}, \bibinfo{person}{Abhishek
  Srivastava}, {and} \bibinfo{person}{Iryna Gurevych}.}
  \bibinfo{year}{2021}\natexlab{a}.
\newblock \showarticletitle{{BEIR}: A Heterogeneous Benchmark for Zero-shot
  Evaluation of Information Retrieval Models}. In
  \bibinfo{booktitle}{\emph{Thirty-fifth Conference on Neural Information
  Processing Systems Datasets and Benchmarks Track (Round 2)}}.
\newblock
\urldef\tempurl%
\url{https://openreview.net/forum?id=wCu6T5xFjeJ}
\showURL{%
\tempurl}


\bibitem[Thakur et~al\mbox{.}(2021b)]%
        {807600ef43073cd9c59d4208ee710e90cf14efa8}
\bibfield{author}{\bibinfo{person}{Nandan Thakur}, \bibinfo{person}{Nils
  Reimers}, \bibinfo{person}{Andreas Ruckl'e}, \bibinfo{person}{Abhishek
  Srivastava}, {and} \bibinfo{person}{Iryna Gurevych}.}
  \bibinfo{year}{2021}\natexlab{b}.
\newblock \bibinfo{title}{BEIR: A Heterogenous Benchmark for Zero-shot
  Evaluation of Information Retrieval Models}.
\newblock


\bibitem[White(2016)]%
        {white2016interactions}
\bibfield{author}{\bibinfo{person}{Robert~W. White}.}
  \bibinfo{year}{2016}\natexlab{}.
\newblock \bibinfo{booktitle}{\emph{Interactions with Search Systems}}.
\newblock \bibinfo{publisher}{Cambridge University Press}.
\newblock


\bibitem[Yao et~al\mbox{.}(2019)]%
        {yao2019empirical}
\bibfield{author}{\bibinfo{person}{Lijun Yao}, \bibinfo{person}{Q. Sun},
  \bibinfo{person}{C. Wang}, {and} \bibinfo{person}{Z. Ding}.}
  \bibinfo{year}{2019}\natexlab{}.
\newblock \showarticletitle{An Empirical Study on Cross-Domain Label Noise}. In
  \bibinfo{booktitle}{\emph{Proceedings of the International Conference on
  Machine Learning}}. \bibinfo{pages}{7068--7077}.
\newblock


\bibitem[Zhan et~al\mbox{.}(2022)]%
        {501397e553ce88c2680c287dc18446e7494ee59d}
\bibfield{author}{\bibinfo{person}{Jingtao Zhan}, \bibinfo{person}{Qingyao Ai},
  \bibinfo{person}{Yiqun Liu}, \bibinfo{person}{Jiaxin Mao},
  \bibinfo{person}{Xiaohui Xie}, \bibinfo{person}{M. Zhang}, {and}
  \bibinfo{person}{Shaoping Ma}.} \bibinfo{year}{2022}\natexlab{}.
\newblock \bibinfo{title}{Disentangled Modeling of Domain and Relevance for
  Adaptable Dense Retrieval}.
\newblock
\href{https://doi.org/10.48550/arXiv.2208.05753}{doi:\nolinkurl{10.48550/arXiv.2208.05753}}


\bibitem[Zhong et~al\mbox{.}(2022)]%
        {Zhong2022ApplyingSA}
\bibfield{author}{\bibinfo{person}{Wei Zhong}, \bibinfo{person}{Yuqing Xie},
  {and} \bibinfo{person}{Jimmy~J. Lin}.} \bibinfo{year}{2022}\natexlab{}.
\newblock \showarticletitle{Applying Structural and Dense Semantic Matching for
  the ARQMath Lab 2022, CLEF}. In \bibinfo{booktitle}{\emph{Conference and Labs
  of the Evaluation Forum}}.
\newblock
\urldef\tempurl%
\url{https://api.semanticscholar.org/CorpusID:251471859}
\showURL{%
\tempurl}


\end{thebibliography}

\appendix
\label{sec:appendix}
\section{Preliminary Experiments}
\label{appendix:preliminary-experiments}

\subsection{Additional Results for IR}
Additional gains in Mean-NDCG are provided for values of $k=[1,5,10,20,30,80]$ on both benchmarks.
Observe that in all regimes fine-tuning on the large FT dataset dramatically improves results for NQ-LLM benchmark.
\begin{figure}[h]
  \centering
  \resizebox{\columnwidth }{!}{%
    \includegraphics{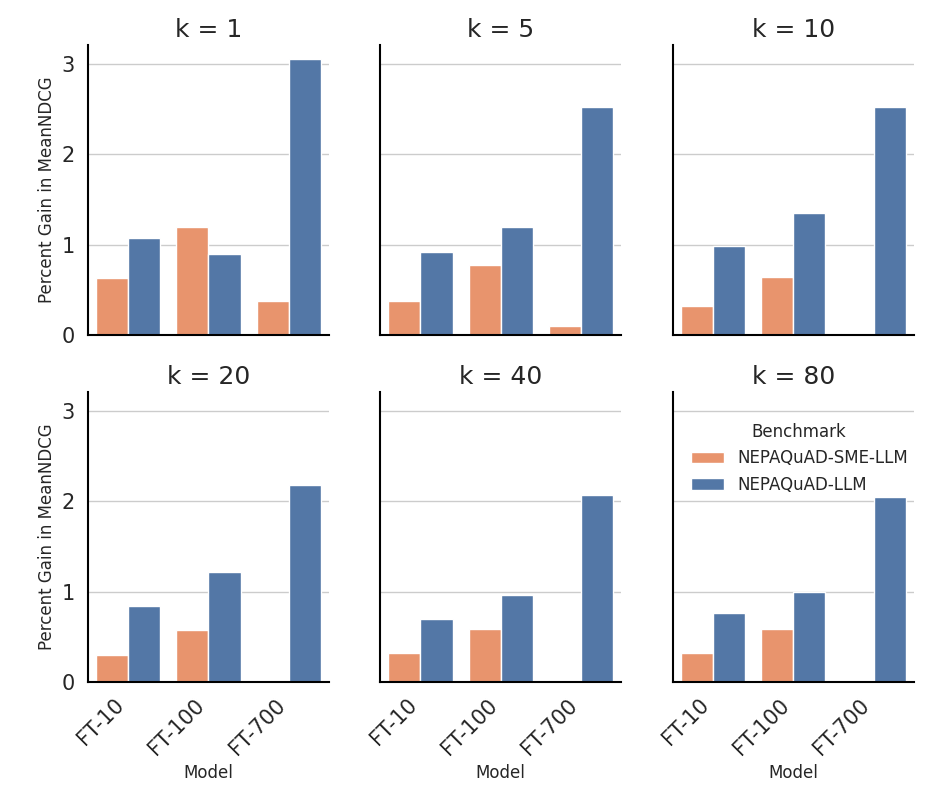}
  }
  \caption{Gain in Mean NDCG for different values of $k=[1,5,10,20,30,80]$ on Nepa-Quad and NQ-LLM.}
  \label{fig:bvk}
\end{figure}

\subsection{Using the BGE Model}
\begin{figure}[h]
  \centering
  \resizebox{\columnwidth}{!}{%
    \includegraphics{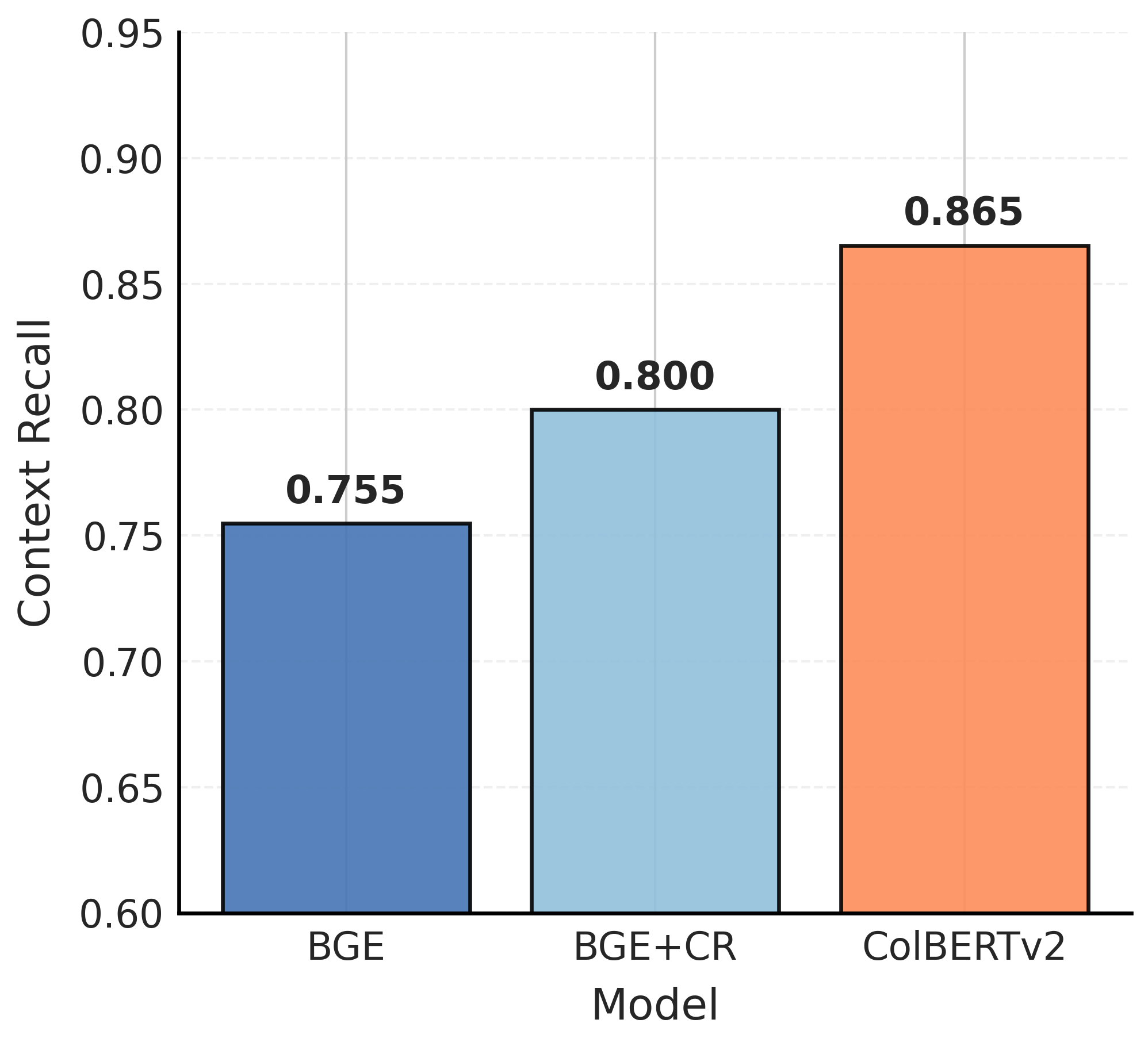}
  }
  \caption{Comparison of Model Performance}
  \label{fig:agg}
\end{figure}

In initial experiments, we used the BAAI/bge-small-en-v1.5 (BGE) embedding model to the NEPA/EIS documents for retrieval tasks. The BGE model, while effective in general-purpose applications, lacked the capability to handle the specialized terminologies and contextual nuances of NEPA documents, resulting in suboptimal retrieval performance (see Figure \ref{fig:agg} for aggregate performance).

\subsection{Combining BGE with ColBERT Reranker}
\begin{figure}[h]
  \centering
  \resizebox{\columnwidth}{!}{%
    \includegraphics{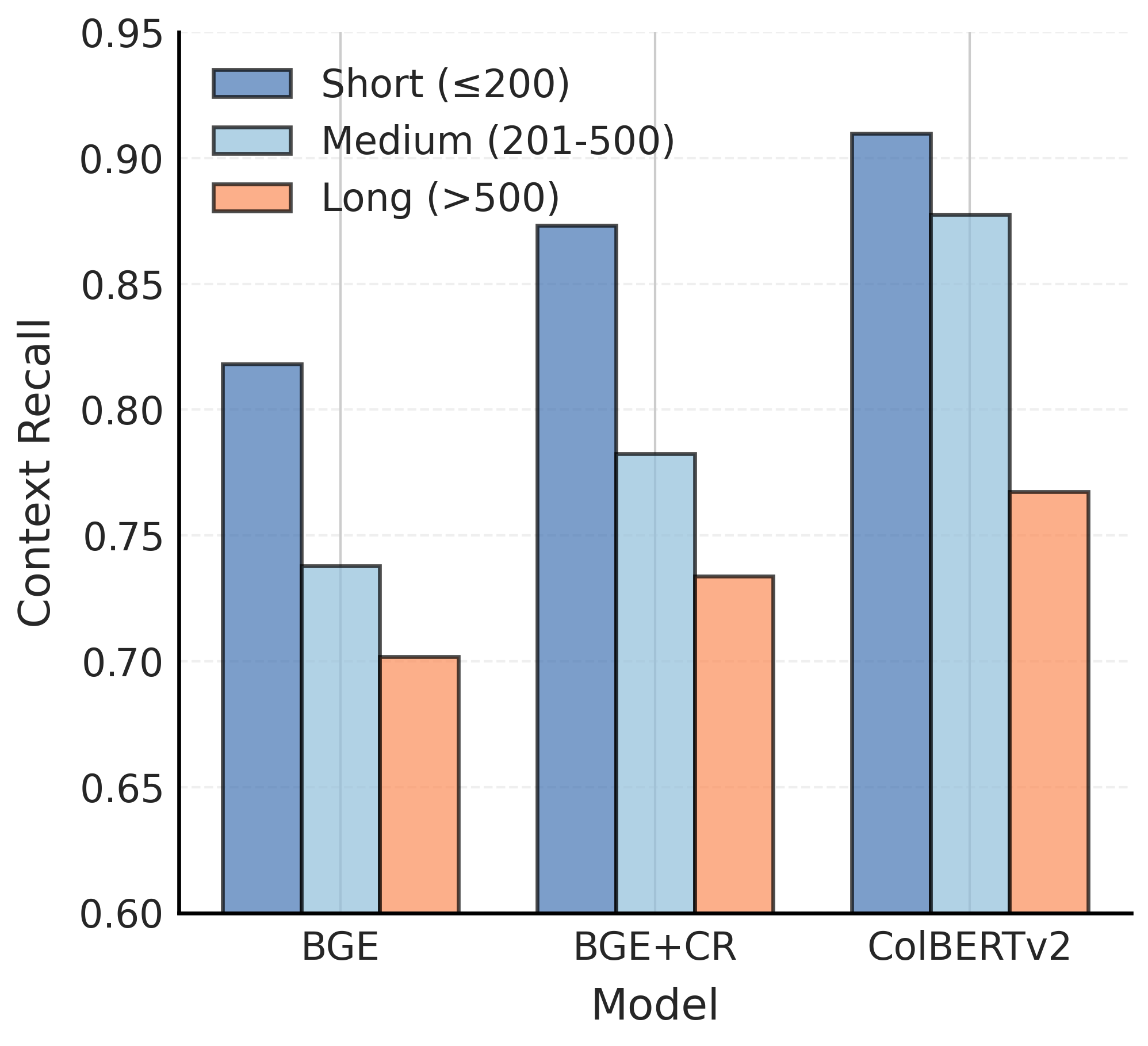}
  }
  \caption{Model Performance by Document Type}
  \label{fig:agg_length}
\end{figure}

To improve performance, we attempted a two-stage retrieval process by using the BGE model for initial retrieval and ColBERTv2 as a reranker. Although this approach produced marginal improvements, it failed to address the fundamental limitations due to the initial embeddings not capturing domain-specific language effectively (see Figure \ref{fig:agg_length} for performance across document lengths).

\subsection{Selection of ColBERTv2 for Adaptation}
These preliminary experiments highlighted the necessity of adopting an embedding model better suited for domain adaptation. ColBERTv2 was chosen due to its ability to perform fine-grained token-level interactions and its demonstrated strength in capturing contextual semantics, making it suitable for adaptation to the NEPA domain.



\section{Training Procedure and Hyperparameters}
\label{appendix:training-procedure}

\subsection{Hardware and Computational Resources}
The model fine tuning was performed on  eight NVIDIA's A100 GPUs.

\subsection{Training Procedure}


Training was conducted using the \texttt{RAGatouille RAGTrainer}, which allowed efficient handling of large datasets and provided tools for optimized training of retrieval-augmented models. The training aimed to minimize a contrastive loss function, enhancing the similarity between queries and their corresponding positive contexts while reducing it with hard negative contexts.

Key aspects of the training procedure included:

\begin{itemize}
    \item Optimization Algorithm: We used the Adam optimizer with appropriate learning rate scheduling to ensure stable convergence during training.
\end{itemize}
\begin{itemize}
    \item Early Stopping: Implemented early stopping criteria based on validation loss to prevent overfitting and promote generalization.
\end{itemize}
\begin{itemize}
    \item Consistency Across Experiments: Maintained the same hyperparameters and training configurations across all experiments, adjusting only the number of training epochs or steps to accommodate the different dataset sizes.
\end{itemize}

\subsection{Training Hyperparameters}

Detailed hyperparameter settings for fine-tuning ColBERTv2 are provided in Table~\ref{table:hyperparameters}. These settings were consistent across all experiments to ensure comparability of results.

\begin{table}[h]
    \centering
    \caption{Hyperparameter for Fine-Tuning ColBERTv2}
    \label{table:hyperparameters}
    \resizebox{\columnwidth}{!}{%
    \begin{tabular}{ll}
        \toprule
        \textbf{Hyperparameter}       & \textbf{Value}        \\
        \midrule
        Batch Size                    & 32                    \\
        Embedding Dimension           & 128                   \\
        Learning Rate                 & $5 \times 10^{-6}$    \\
        Maximum Sequence Length       & 256 tokens            \\
        Number of Training Epochs     & Adjusted per dataset size  \\
        Optimizer                     & Adam                  \\
        Early Stopping Criteria       & Validation loss       \\
        Loss Function                 & Contrastive Loss      \\
        \bottomrule
    \end{tabular}%
    }
\end{table}

\subsection{Software and Tools}

Programming Language : Python; Deep Learning Framework : PyTorch; Training Framework : \texttt{RAGatouille RAGTrainer}; Language Models : \texttt{Gemini 1.5 Pro}, ColBERTv2 ; Libraries : Text segmentation tools, other Python libraries for data handling and processing

\section{NEPAQuAD-LLM}
\label{appendix:test-dataset}

\subsection{Question Types}

The NEPAQuAD-LLM (NQ-LLM) benchmark includes a variety of question types to reflect actual information needs within the NEPA domain. The distribution of question types is as follows:

\begin{table}[htbp]
  \centering
  \caption{Question Types and Counts}
  \label{tab:question_types}
  \begin{tabular}{lr}
    \toprule
    \textbf{Question Type} & \textbf{Count} \\
    \midrule
    Inference      & 287 \\
    Closed-ended   & 148 \\
    Comparison     & 50  \\
    Process        & 35  \\
    Divergent      & 21  \\
    Evaluation     & 15  \\
    \bottomrule
  \end{tabular}
\end{table}

\subsection{Document Sources}

The NQ-LLM benchmark encompasses queries from ten EIS documents, ensuring diversity in content and agency representation. Each document contributed between 51 to 63 questions to the test set. The documents include:

\begin{table}[htbp]
  \centering
  \caption{Questions Generated from Each EIS Document File}
  \label{tab:queries_per_file}
  \begin{tabular}{lr}
    \toprule
    \textbf{Document Source} & \textbf{\# Questions} \\
    \midrule
    Goldrush Mine Project FEIS & 63 \\
    Continental US Interceptor Site & 63 \\
    Final Tank Closure & 57 \\
    Fort Wainwright Alaska & 57 \\
    Alaska LNG Project & 56 \\
    Land Management Plan & 55 \\
    T7A Recapitalization & 52 \\
    Sea Port Oil Terminal & 51 \\
    FirstNet & 51 \\
    PEIS for Oil and Gas & 51 \\
    \bottomrule
  \end{tabular}
\end{table}

\section{Agency Distribution in Evaluation Benchmarks}
\label{appendix:agency_distribution}

Table~\ref{table:agency_distribution} shows the complete breakdown of federal agencies that originated or contributed to the Environmental Impact Statements in our evaluation benchmarks. A total of 12 federal agencies contributed to these EIS documents, with multiple agencies sometimes contributing to individual EIS document. Both NEPAQuAD-SME-LLM and NEPAQuAD-LLM datasets ensure diverse representation of regulatory contexts and terminology while maintaining comparable agency coverage between benchmarks.

\begin{table}[htbp]
  \centering
  \caption{Federal agencies represented in evaluation benchmarks}
  \label{table:agency_distribution}
  \begin{tabular}{lr}
    \toprule
    \textbf{Federal Agency} & \textbf{\# Docs} \\
    \midrule
    U.S. Department of Energy (DOE) & 2 \\
    Missile Defense Agency (MDA) & 1 \\
    U.S. Department of Commerce & 1 \\
    U.S. Army Garrison Alaska & 1 \\
    Bureau of Land Management & 1 \\
    USDA Forest Service & 1 \\
    Bureau of Safety and Environmental Enforcement & 1 \\
    Bureau of Ocean Energy Management & 1 \\
    U.S. Coast Guard (USCG) & 1 \\
    Maritime Administration (MARAD) & 1 \\
    U.S. Department of the Air Force & 1 \\
    Air Education and Training Command & 1 \\
    \bottomrule
  \end{tabular}
\end{table}

\section{NQ-LLM Benchmark Generation}
\label{appendix:prompt}
Below is the prompt we used to generate question-context pair for the NQ-LLM Benchmark. 


\begin{tcolorbox}[colback=gray!5!white, colframe=gray!75!black, title=Prompt]

\textit{You are an advanced AI system with expertise in natural language processing and question generation. Your task is to assist in creating a high-quality, diverse synthetic dataset for training information retrieval models.}

\vspace{0.5em}
\textbf{Given the entire report below, perform the following steps:}

\begin{enumerate}
    \item Carefully read and analyze the report to understand its content, main ideas, and key details.
    
    \item Generate thought-provoking questions based on the content of the report, along with their corresponding contexts. For each pair:
    
    \begin{itemize}
        \item Select a relevant context from the report that is 3-4 lines long and provides a comprehensive picture to answer the question without requiring external knowledge.
        
        \item Generate a question that is directly relevant to the selected context.
        
        \item The question should cover one of the following types:
        \begin{itemize}
            \item \textbf{Closed-ended:} Questions that can be answered with a simple 'yes' or 'no' based on the information provided in the context.
            
            \item \textbf{Comparison:} Questions that require comparing and contrasting information from the context, involving similarities, differences, or temporal changes.
            
            \item \textbf{Divergent:} Open-ended questions that require using information from the context to extrapolate, infer, or explore possibilities.
            
            \item \textbf{Evaluation:} Questions that ask for an assessment or judgment based on the information in the context.
            
            \item \textbf{Inference:} Questions that require reading between the lines and drawing conclusions based on the information provided.
            
            \item \textbf{Process:} Questions that ask about how something works or the steps involved in a process described in the context.
        \end{itemize}
        
        \item Ensure that each question is concise, clear, and grammatically correct.
        
        \item Confirm that the selected context contains all the necessary details to answer the generated question. The answer should be directly derivable from the given context without requiring external knowledge.
    \end{itemize}
    
    \item Provide the generated question-context pairs.
\end{enumerate}

\vspace{0.5em}
\textbf{Remember:} The goal is to create a diverse set of challenging questions that effectively test the model's ability to retrieve and understand relevant information from the given report. Maintain high-quality standards throughout the dataset generation process.

\end{tcolorbox}

\end{document}